# In Situ Characterisation of Graphene Growth on Liquid Copper-Gallium Alloys: Paving the Path for Cost-Effective Synthesis

Valentina Rein[1,*], Florian Letellier[1], Maciej Jankowski[1], Marc de Voogd[2], Mahesh Prabhu[3], Lipeng Yao[1], Gertjan van Baarle[2], Gilles Renaud[4], Mehdi Saedi[5], Irene M. N. Groot[3], Oleg V. Konovalov[1,*]

[1] *European Synchrotron Radiation Facility - ESRF, 71 Avenue des Martyrs, 38043 Grenoble, France*

[2] *Leiden Probe Microscopy (LPM), Kenauweg 21, 2331 BA Leiden, The Netherlands*

[3] *Leiden Institute of Chemistry, Leiden University, P.O. Box 9502, 2300 RA Leiden, The Netherlands*

[4] *Univ. Grenoble Alpes and CEA, IRIG/MEM/NRS, 38000 Grenoble, France*

[5] *Physics Department, Shahid Beheshti University, 1983969411 Tehran, Iran*

\* valentina.belova@esrf.fr; konovalo@esrf.fr

**ABSTRACT**

Liquid metal catalysts (LMCats), primarily molten copper, have demonstrated their efficiency in the chemical vapour deposition (CVD) approach for synthesising high-quality, large-area graphene. However, their high melting temperatures limit broader applications. Reducing the temperature of graphene production on LMCats would lead to a more efficient and cost-effective process. Here, we investigated the effects of alloying copper with a low-melting temperature metal on graphene growth in real-time. We examined a set of liquid copper-gallium alloy systems using two complementary in situ techniques: radiation-mode optical microscopy and synchrotron X-ray reflectivity (XRR). Microscopy observations revealed reduced catalytic activity and graphene quality degradation in compositions with gallium domination. The XRR confirmed the formation of single-layer graphene on alloys with up to 60 wt% of gallium. Additionally, we detected a systematic increase in adsorption height on the alloys' surface, suggesting a weaker graphene adhesion on gallium. These findings propose a trade-off between layer quality and production cost reduction is feasible. Our results offer insights into the CVD synthesis of graphene on bimetallic liquid surfaces and underscore the potential of gallium-copper alloys for enabling the direct transfer of graphene from a liquid substrate, thereby addressing the limitations imposed by high melting temperatures of conventional LMCats.

# 1. Introduction

The unique properties and versatility of two-dimensional materials (2DMs) such as graphene, germanene, h-BN, transition metal dichalcogenides, and others, along with their potential for industrial production, make them an exciting area of research with significant options for technological advancement [1,2]. Graphene was the first one to be isolated in 2004 and has become the benchmark of the 2DM family [3]. Since then, extensive studies helped to gain a deep understanding of graphene's unique properties, including high mechanical strength, electrical conductivity, thermal conductivity, flexibility, and transparency [4]. However, these promising properties of graphene largely depend on its quality, which is defined by its crystallinity and defect density.

Over the years, the development of methods to elaborate high-quality graphene on a large scale has been an active area of research. The standard method for synthesising large-area graphene is chemical vapour deposition (CVD), which is commonly based on the catalytic decomposition of a hydrocarbon precursor gas on a solid metal substrate [5]. Based on experimental evidence, copper is considered the most effective substrate for graphene growth. The low solubility of carbon in Cu enables the growth of single-layer graphene over large areas [6]. Despite its wide application, several shortcomings associated with CVD on a solid substrate significantly compromise the quality of the graphene produced [7]. The solid substrate induces defects such as grain boundaries, vacancies, and lattice distortions. The growth rate and uniformity of the graphene sheet are hard to control, resulting in significant variations in its thickness and quality across the substrate. Transferring the graphene from the solid substrate to another support inflicts further layer degradation. Ongoing research focuses on improving CVD-fabricated graphene's scalability, yield, and quality. Developing new techniques to overcome the limitations of this method is critical for the advanced applications and commercialisation of graphene and other 2DMs in highly demanding domains such as micro- and nanotechnology.

The groundbreaking idea to replace solid metal catalysts with their liquid counterparts (LMCats) was first proposed by Wu *et al*. [8]. This approach yields high-quality, large-area single-crystal 2DMs with fewer defects and impurities [9,10]. LMCat substrates such as liquid Cu have an atomically smooth surface, yielding a low density of nucleation centres and fast mass transport, allowing for high growth rates that significantly reduce the density of defects and domain boundaries [11,12]. However, studies of LMCat-based CVD systems typically require cooling down to room temperature, thus re-solidifying the substrate and significantly altering the grown graphene. In addition, the detailed information about the growth dynamics remained unknown.

Recently, we developed a customised CVD reactor, adapted to harsh conditions of CVD on LMCats, *e.g.* liquid Cu: a temperature of ~1400 K, a pressure of 200 mbar including CVD gas precursors, and intense metal evaporation. The reactor allows for X-ray reflectivity (XRR) and other X-ray scattering techniques, Raman spectroscopy, and radiation-mode optical microscopy to be combined for *in situ* characterisation in real-time during the growth [13]. The setup allows for precise monitoring and control of graphene growth on liquid Cu in real time with high reproducibility [11,12]. The high quality of the synthesised graphene was confirmed by Raman characterisation.

Solidification of the liquid substrate would result in the deterioration of the as-grown graphene quality. Therefore, the ultimate goal is a direct separation of continuously produced graphene from the hot liquid substrate and then its seamless transfer to target support. The successful development of such direct separation method will substantially depend on the physical and chemical interactions between the graphene and the LMCat, such as adsorption energy, catalytic activity, and wettability, as well as the mechanical stability of the layer. The crucial parameter here is metal-graphene interactions, which allow for applying the most feasible transfer method to separate graphene from the substrate [14]. Getting insight into these interactions demands understanding the interface's structure and proper modelling, which is well-established for solids [15–18] and supported by some experimental examples [19–22] but very scarce for liquid surfaces [23]. Furthermore, the fracture strength of graphene is predicted to decrease with a temperature rise [24]. Also, at high temperatures, the separated graphene has an enhanced risk of reacting with hydrogen or gas impurities (*e.g.* oxygen) [25,26]. Therefore, using low-melting-point LMCats to reduce the transfer process's temperature without solidifying the substrate is advantageous. The growth of graphene on Ga with a melting point of 303 K has already been reported in the literature [27–30]. However, these studies were conducted *ex situ* and post-sample solidification, leaving the detailed mechanisms of graphene formation largely unexplored. Given gallium's low melting point and its low adhesion energy with graphene [31], we consider it to be among the most promising LMCats for developing direct transfer technologies.

In addition, using lower temperatures during graphene growth will reduce energy consumption. Besides the obvious economic benefits, it will simplify designing and operating the reactor for 2DM synthesis and separation. However, synthesising graphene at lower temperatures may also impede its growth rate and increase the number of defects. Hence, the conditions for graphene growth on Ga, which would allow a reasonable compromise between the quality and the gain in process temperature reduction, should be investigated. For this purpose, we explored *in situ* the growth of graphene on liquid Ga and CuGa alloys of different compositions and, thus, melting points. A series of CuGa samples with varying compositions ensures a gradual transition from the well-studied copper system to the new one of gallium. Comparing the results with those from Cu is essential in defining the most efficient growth process. The results obtained and the acquired technical experience will help investigate the direct separation process using low-melting-point LMCats.

We used the LMCat reactor [13] equipped with a quartz window port filled with flowing methane as a precursor in a mixture of argon and hydrogen. The substrate temperature $T_s$ was varied between an upper value of 1463 K and the melting point $T_m$ of the corresponding alloy: ~1360, ~1300, ~1210, ~1140, ~1070, ~970, ~870, ~780, and ~300 K for pure Cu, 10%, 20%, 30%, 40%, 50%, 60%, 70% alloys, and pure Ga, respectively [32]. Two complementary *in situ* methods were employed: radiation-mode optical microscopy and XRR. *In situ* optical microscopy provides information on the nucleation, growth rate, movements, and interactions of 2DMs on LMCats, while XRR helps to characterise the grown layers' structural properties.

## 2. Results and discussion

### *2.1. Radiation-mode optical microscopy*

Our recent work explored graphene's growth behaviour on liquid Cu [11,12]. Under our typical experimental conditions, the growth starts with nucleation and formation of separate flakes whose shapes vary between circular, hexagonal, and concave-dodecagonal depending on the partial pressures *p* of the CVD precursors $CH_4$ and $H_2$. Then, due to an interplay between repulsive and attractive forces, the flakes self-assemble in a hexagonal lattice, filling the remaining surface of the liquid metal pool, followed by their coalescence in the last growth stage. Before coalescing, flakes can reach millimetres in lateral size depending on the growth rates. The remarkably low defect density of these flakes was proven by Raman spectroscopy characterisation.

To investigate whether and how Ga modifies the growth mechanism(s) of graphene and the catalytic activity of the liquid substrate, we prepared a series of $Cu_{(1-x)}Ga_x$ alloy samples with varying Ga weight composition *x* from 10 to 100%. The microscopy images are obtained through the quartz window port above the sample thanks to the difference in the thermal emissivity between the graphene layer and the liquid metal. However, the measured emissivity of CuGa alloys is higher than those of the pure metals separately, as illustrated in the Supporting Information (SI), Fig. S1. It leads to decreased emissivity contrast between the graphene flakes and the substrate for those samples.

In general, the formation of carbon structures was observed for the entire concentration range of CuGa alloys. Fig. 1 shows the optical microscopy images of the typical growth modes observed for different coverage stages during the growth as a function of Ga composition. The corresponding video files of the growths are accessible online as Movies S1-S5. The growth mechanism undergoes significant transformations from pure Cu [11,12] to pure Ga. No drastic change in the growth mechanism is observed after adding 10% of Ga to pure Cu (Fig. 1a–c). However, the growth rates and average flake sizes are smaller. Mostly, only the circular shape is seen. Also, the merging of the flakes is observed at earlier growth stages, leading to the formation of branch-like structures. It may indicate a disturbance of the balance of the repulsion-attraction interdomain forces toward a more substantial attractive contribution.

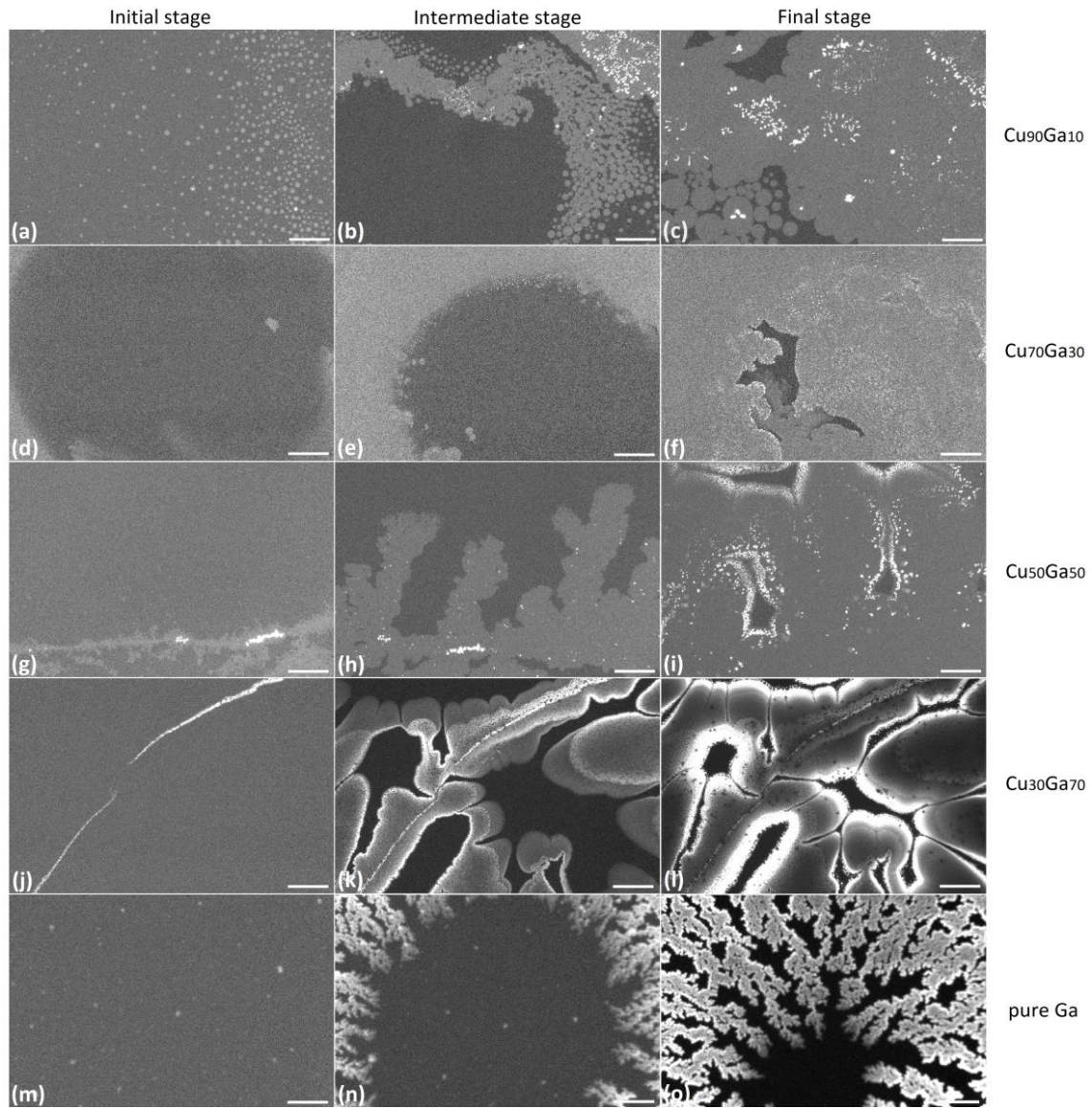

**Fig. 1.** Evolution of the growth mechanism of carbon layers on a liquid surface of CuGa alloys with an increase of Ga weight percentage from top to bottom. Different coverage stages are shown from left to right for the samples of 10, 30, 50, 70, and 100% of Ga (the intermediate concentrations of 20, 40, and 60% were also studied but omitted from the figure for clarity). The scale bar is 0.4 mm. The intensity scale corresponds to the emissivity ratio and is the same for all images. The darkest contrast of the colour scale corresponds to the liquid metal surface, and the brightness scales linearly with the number of carbon layers. The corresponding video files are accessible online as Movies S1-S5.

Further increase of the Ga content results in a further decrease in flake sizes. As deduced from Fig. 1d–f, for the $Cu_{70}Ga_{30}$ sample, most flakes are smaller than the camera's resolution, which is a few micrometres. Therefore, the assemblies of flakes appear as clouds due to a contrast gradient from the areas of highly concentrated flakes (light shade) to the clean alloy surface (dark shade). The spreading of the graphene layer may occur either through the formation of branches or through a flat front propagation from the sample edges toward the centre. Noteworthy, the growth slows down with higher coverage.

In the samples with a higher percentage of Ga, it is impossible to distinguish single-crystal flakes anymore. However, in the early stages, we can sometimes observe flakes of irregular shapes that seem to be clusters of graphene domains (Fig. 1g, m, n). As can be seen in the examples of 50 (Fig. 1g–i), 70 (Fig. 1j–l), and 100% of Ga (Fig. 1m–o), the tendency to anisotropic growth becomes stronger: the branches get narrower, giving rise to a more significant number of sprouts. Gallium also seems to facilitate the formation of multilayers and 3D structures, having high intensity on the images due to the linear dependence of the emissivity on the number of carbon layers [11]. Consequently, we observe a distinct transition of the growth regime from forming individual single crystals of graphene with isotropic 2D expansion to strongly anisotropic dendritic growth of carbon 3D structures.

Next, we collected flakes' size and number statistics for the growth stage preceding the coalescence, presented in the SI, Fig. S2. The results of the analysis of the growth rates as a function of $p_{CH4}/p_{H2}$ and $T_s$ are demonstrated in Fig. 2. We see a notable decrease by more than one order of magnitude of the flakes' growth rates (open symbols) upon admixture of 10 and 20% of Ga compared with pure Cu (Fig. 2a–c). For the low-Ga contents, when individual flakes are distinguishable, the growth rate is that of the average flake radius of a disk having the flake area. For the samples with a higher content of Ga (≥30%), we do not detect separate graphene flakes and instead use the growth rate of the edge of the growing graphene layer (filled symbols). Starting from the most minor Ga concentration considered in our study, *i.e.*, $Cu_{90}Ga_{10}$, we notice a different growth mode: a continuous propagation of the layer from the LMCat edges with a pretty high rate (>20 μm s$^{-1}$). Counterintuitively, such behaviour seems to be activated by lowering $T_s$ close to the melting point $T_m$ of Cu or even lower (1363 K and 1308 K in Fig. 2b). The same trend of layer propagation at low $T_s$ (below copper $T_m$ of 1358 K) with rates higher or similar to those at higher $T_s$ (above the $T_m$ of Cu) is observed up to 50% of Ga, although below 1200 K the rates drastically decrease. For 20% Ga, we see that the flake radius and edge growth rates match well. Further increase in the concentration of Ga does not lead to a significant drop in the detectable growth rates—however, the $p_{CH4}/p_{H2}$ threshold for the graphene growth shifts towards larger values. Thus, when moving from a 30% sample to pure Ga, it is necessary to increase $p_{CH4}$ by two orders of magnitude to initiate and/or continue the growth (Fig. 2d–i). Together with the reduced growth rates and layer quality, it corroborates the substantial decrease in the catalytic activity of Cu upon alloying with Ga and the inferior catalytic activity of Ga with respect to Cu. The differences in the growth and the transport of C species may be related to the high carbon solubility within an ultra-thin surface layer of Ga suggested by Ueki *et al.* [33].

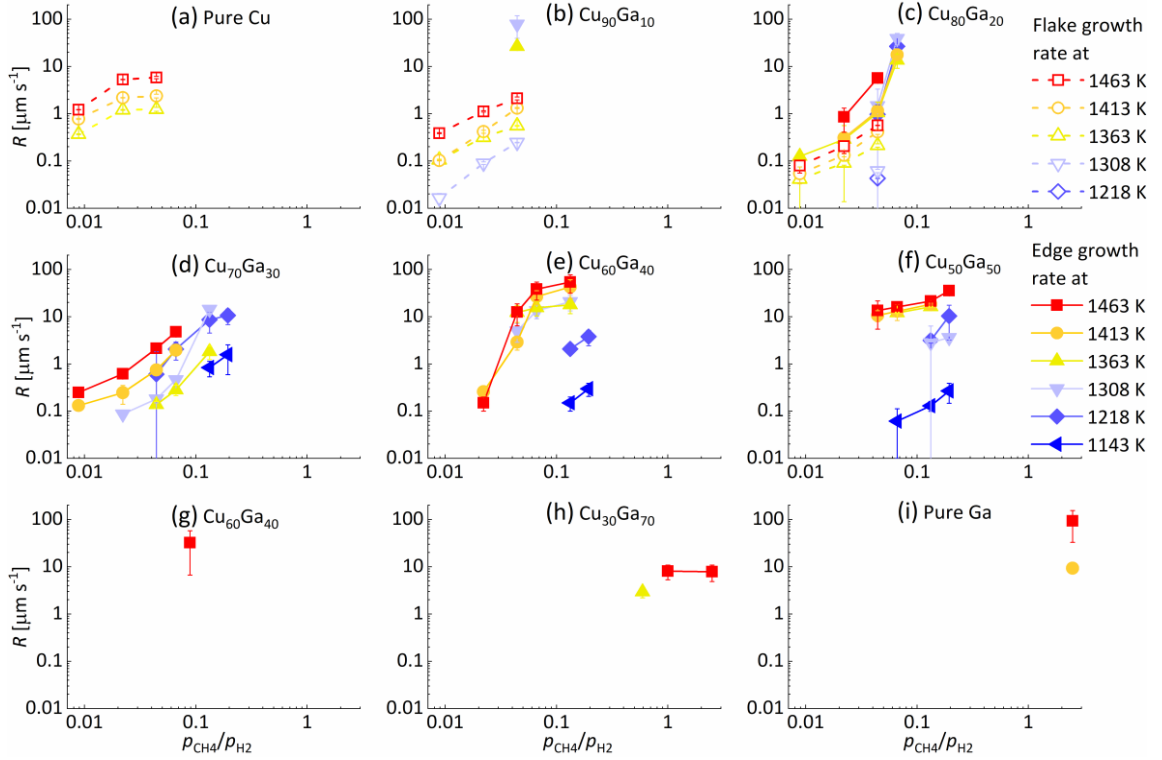

**Fig. 2.** Flake radius growth rate $R$ (open symbols) and edge growth rate (filled symbols) as a function of $p$ and $T_s$ (colour-coded) for the series of CuGa alloy compositions. A log-log scale is used.

## 2.2. X-ray reflectivity

Using the synchrotron XRR, we performed an *in situ* study of graphene layers grown *via* CVD on liquid CuGa alloys. We recorded XRR curves on the clean liquid alloy surfaces before and after the graphene growth. The reflectivity data from the curved liquid surfaces were treated according to the procedure described in Ref. [34]. Due to the sample curvature, the beam footprint on the surface varies between hundreds of microns, providing sufficient areal averaging [35]. The resulting curves are plotted (open symbols) in Fig. 3a along with the fit (red solid lines), aiming to extract layers' roughness, thickness, and electron density profiles. The results of the fit, presented in Table 1, will be discussed further below.

As shown in the previous section, adding a small amount of Ga does not drastically change the growth mechanism. The XRR curves from graphene layers on alloys up to 60 wt% of Ga resemble that of graphene on pure Cu (the upper curve). The presence of only two Kiessig oscillations in the measured range, with a pronounced minimum for the scattering vector $q_z$ in the range of 0.7–0.9 Å$^{-1}$ and the absence of a Bragg peak are the clear signatures of single-layer graphene [36]. The low roughness is also evidence of the reasonably good quality of the grown layers.

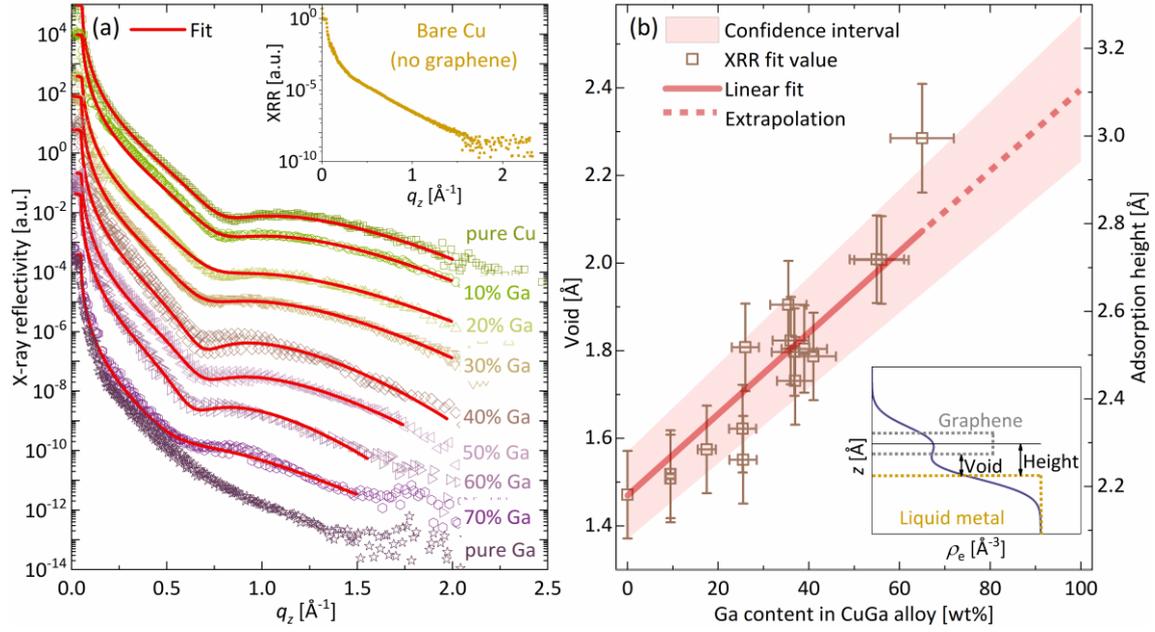

**Fig. 3.** (a) XRR curves of the series of liquid CuGa alloy samples with CVD-grown graphene/graphite on top and the fit based on a slab model. The curves are vertically offset for clarity. The inset shows the XRR curve of bare liquid metal. (b) Graphene-copper adsorption height extracted from the fit of the XRR curves (void between Cu and C slabs plus half of the C slab, 0.71 Å). Single-layer graphene could not be observed and measured on pure Ga; therefore, no height value is reported. The inset illustrates the '*void*' and '*height*' definitions. The alloy composition deviates from the nominal ones due to an intense Ga evaporation in the liquid state. Thus, the horizontal error bar increases with the Ga contribution. The vertical error bar of 0.10 Å is based on the experimental reproducibility throughout the previous studies. The linear fit with a slope of 0.0093 ± 0.0007 Å/% demonstrates the linear dependence of the graphene-metal distance on the content of Ga. The confidence interval is a combination of these two uncertainties.

For the 70 and 100% Ga samples, the peak around 1.8 Å$^{-1}$, which is close to the (002) Bragg peak of graphite, agrees with the change in growth mode above 60 wt% of Ga to dendritic and 3D growth [37]. Note that these multilayers could not be etched at variance with single-layer graphene by increasing the H$_2$/CH$_4$ partial pressure ratio. The difficulty in achieving single-layer graphene on pristine Ga does not come as a surprise and could be attributed to variations in experimental conditions compared to previous attempts reported in the literature. On the one hand, gallium's ability to promote the formation of multilayer graphene and graphitic structures was observed previously [27,28,33,38–40]. On the other hand, the earlier reports on graphene CVD growth on liquid Ga did not include the use of H$_2$. Consequently, recent theoretical work [41] casts doubt on previous results on graphene grown on Ga. It suggests that the reported attempts [27-30] were likely conducted under a residual pressure of O$_2$, which typically reacts with the Ga surface to form an oxide skin under the described conditions. Our observations also confirm that removing the oxide layer from liquid Ga requires etching at high temperatures in the H₂ atmosphere. Since, in our experiment, the oxide layer was etched away prior to carbon deposition, and H₂ was constantly flowing during the CVD process, we can safely exclude the above-mentioned side effect.

Another interesting observation in the XRR curves is the gradual shift of the first Kiessig minimum toward low $q_z$ with increasing Ga content. We used the Refl1d software to fit the XRR curves [42], applying a slab model containing three slabs: Cu, C, and a separation void in between. The densities of liquid Cu and liquid Ga were taken from the literature [43,44]. The corresponding free parameters of the fit are summarised in Table 1. We recorded multiple scans for some compositions, the analysis of which can be found in the SI, Table S1. The void thickness directly corresponds to an important parameter – the interlayer graphene-copper distance typically attributed to van der Waals interaction, which we also denote as '*adsorption height*' [23]. In our previous works, we define this parameter as the distance between the inflexion point of the Cu electron density $\rho_e$ profile at the interface and the centre of the carbon slab [11,23,34,35]. Thus, the height is composed of the thickness of the void and half of the graphene layer thickness (1.42/2 Å). The height values obtained by fitting the XRR curves for the series of CuGa samples are shown in Fig. 3b, where we can observe a systematic increase with increasing Ga percentage. Due to intense metal evaporation, the actual content of the CuGa samples shifts towards lower Ga contribution, as confirmed by post-growth scanning electron microscopy (SEM) with energy-dispersive X-ray analysis (EDX) (see the SI for more details, Tables S2, S3, and Fig. S3). Based on the EDX results, we set the uncertainty as 10% of the nominal Ga percentage. We apply an error bar of 0.10 Å for the height values based on the experimental reproducibility according to our previous works [11,23,34,35]. Assuming a linear dependence of the distance with Ga concentration (the slope resulting from the linear fit is 0.0093 ± 0.0007 Å/%), the extrapolation suggests an increase from 2.18 ± 0.10 Å on liquid Cu to 3.11 ± 0.17 Å on liquid Ga. To estimate the confidence interval, we combine the error bar of the starting point of 0.10 Å and the fit uncertainty for the slope.

Qualitatively, the value of the adsorption height allows for estimating the strength of the graphene-metal bonding and the work needed to separate the graphene from its substrate, *e.g.*, to transfer it to another substrate. This trend is now well-documented for solid, crystalline, and transition metal substrates. Strong bonding, *e.g.*, with Co, Ni, Ru, Rh, and Re, typically corresponds to graphene-metal distances of ~2.1 Å. In contrast, weak bonding (*e.g.*, with Cu, Ag, Al, Ir, Pt, and Au) corresponds to more than 1 Å larger distances of ~3–3.4 Å, typical of weak van der Waals bonding (close to the interlayer distance of 3.35 Å in bulk graphite) [18]. In a first rough approximation [45], the bonding of graphene with metals is primarily due to the metal d-band electrons. Indeed, Batzill *et al.* [45] observed that the metal-graphene distance increases when the distance between the *d*-band and the Fermi level increases. Thus, we expect that the metal distance increase from Co to Ni, then to Cu, shall continue to increase for the following metals, Zn and Ga. Note that, strictly speaking, Zn and Ga are not transition metals since their $3d$ band is filled, like for Cu. However, we can expect that the further filling of the $4s$ and $4p$ bands will only further weaken the bond strength. Another factor along this trend could be the larger van der Waals radius of Ga (1.87 Å) compared to Cu (1.40 Å). Ga can also be considered electronically similar to Al, which is just on top of the same row (13th) of the periodic table of the elements. The graphene–to–Al distance is the highest reported (3.41 Å).

**Table 1.** XRR fit results: the roughness of the liquid metal surface $\sigma_m$, the roughness of the carbon layer $\sigma_C$, and the thickness of the separation void $t_v$. The error bars in the last three columns result from the fitting procedure.

| Nominal Ga content, wt% | Estimated actual Ga content, wt%[*] | $\sigma_m$, Å | $\sigma_C$, Å | $t_v$, Å |
|---|---|---|---|---|
| 0 | 0 | 1.099 ± 0.008 | 1.119 ± 0.004 | 1.471 ± 0.007 |
| 10 | 9.5 ± 1.0 | 1.091 ± 0.013 | 1.178 ± 0.010 | 1.509 ± 0.017 |
| 20 | 17.5 ± 2.0 | 1.009 ± 0.019 | 1.194 ± 0.022 | 1.575 ± 0.028 |
| 30 | 25.4 ± 3.0 | 0.838 ± 0.028 | 1.162 ± 0.015 | 1.622 ± 0.022 |
| 40 | 36.0 ± 4.0 | 1.432 ± 0.024 | 1.422 ± 0.014 | 1.823 ± 0.024 |
| 50 | 39.0 ± 5.0 | 1.224 ± 0.008 | 1.599 ± 0.010 | 1.804 ± 0.008 |
| 60 | 56.0 ± 6.0 | 1.510 ± 0.015 | 1.626 ± 0.011 | 2.007 ± 0.016 |
| 70 | 65.0 ± 7.0 | 0.904 ± 0.050 | 2.076 ± 0.080 | 2.285 ± 0.124 |

[*]at the moment of scanning

## 3. Conclusion

In summary, this study investigates the effect of copper-gallium alloying on the growth kinetics of graphene and assesses the potential of CuGa alloys as a liquid catalyst for cost-effective graphene synthesis by CVD. It is done by combining *in situ* radiation-mode optical microscopy and XRR to track the growth mode of graphene on the liquid substrate in real time. We find that adding Ga to Cu drastically alters the growth mechanism of graphene: the growth evolves from fast to slow, and the grown material undergoes a transition from large separate flakes to continuous growth, which then evolves into slow branched 3D growth as Ga dominates. Furthermore, the XRR data shows the formation of single-layer graphene up to 60 wt% of Ga (with the melting point around 873 K *vs* 1358 K of Cu) and a gradual shift of the first Kiessig minimum in the direction of low $q_z$ with increasing Ga content, which points toward a somewhat weaker van der Waals interaction between graphene and gallium in comparison to copper.

The findings of this study provide insight into the synthesis of graphene and highlight the potential of CuGa alloys to serve as liquid metal catalysts with a weaker adhesion to graphene and a melting temperature reduced by a few hundred degrees as a replacement for high-melting temperature copper. Furthermore, this is achieved while maintaining a decent catalytic activity and the quality of the grown graphene in the case of the alloys with medium Ga content. The reported observations are expected to facilitate the development of technologies for the direct separation of graphene from the liquid substrate without solidification and degradation of the quality of the grown layer. Therefore, our research has significant potential for implications in the production technologies and commercialisation of graphene and other 2DMs. The experimentally obtained parameters are also of high interest for theoretical studies that often lack the experimental input in the domain of molecular simulations.

## 4. Materials and methods

In order to investigate the catalytic activity of liquid Ga for CVD graphene growth compared to liquid Cu, we prepared a series of CuGa alloys with varying content of Ga from 10 wt% to pure Ga. The Cu foils of 99.9976% purity were purchased from Advent Research Materials (Eynsham, The United Kingdom), and the Ga lump of 99.9999% purity was purchased from Goodfellow (Lille, France), respectively. The alloy samples were prepared by melting Cu and Ga pieces on a tungsten disk as a sample holder inside the reactor prior to the experiments. The tungsten disks with a diameter of 25 mm were purchased from Metel BV (Waalwijk, The Netherlands). The percentage of Ga was controlled by the change of the melting point $T_m$ according to the known phase diagram [32]. The mobile CVD reactor with the attached gas system was designed particularly for *in situ* optical and X-ray measurements of graphene growth at high temperatures above liquid Cu [13]. The growth was monitored and recorded using optical microscopy in the radiation mode with a ×5 objective and a digital camera. Due to the fact that the emissivity of graphene is somewhat higher than that of liquid Cu, the flakes and layers of graphene can be visually distinguished by higher contrast. Moreover, the brightness scales linearly with the number of carbon layers [11]. The experimental range of $T_s$ was limited by the solidification point of the alloys on the lower side to a maximum of 1473 K restricted by the heater power and safety. A standard set of gases purchased from Air Liquide (Paris, France) was employed: Ar, $H_2$, and $CH_4$. The methane gas was delivered in bottles of two different concentrations: 5% diluted in Ar and 100%. The flowmeters built into the gas system controlled the partial pressure ratio of $CH_4$ to $H_2$ and varied between 0 and 2.5.

The experimental conditions for the growths shown in Fig. 1 in the main text and online video files Movies S1–S5 were:

$Cu_{90}Ga_{10}$ (Movie S1): $T_s$ = 1413 K, $P_{CH4}/P_{H2}$ = 4.4×10$^{-2}$

$Cu_{70}Ga_{30}$ (Movie S2): $T_s$ = 1218 K, $P_{CH4}/P_{H2}$ = 1.3×10$^{-1}$-1.9×10$^{-1}$

$Cu_{50}Ga_{50}$ (Movie S3): $T_s$ = 1463 K, $P_{CH4}/P_{H2}$ = 4.4×10$^{-2}$-1.3×10$^{-1}$

Cu$_{30}$Ga$_{70}$ (Movie S4): $T_s$ = 1463 K, $P_{CH4}/P_{H2}$ = 1.0-2.5

Ga$_{100}$ (Movie S5): $T_s$ = 1463 K, $P_{CH4}/P_{H2}$ = 2.5

The X-ray study was carried out at beamline ID10 of the European Synchrotron Radiation Facilities (ESRF, France). The beamline is equipped with a double-crystal deflector that allows defining and varying the incident angle of the X-ray beam with respect to liquid surfaces while keeping them horizontal [46]. The beam energy was set to 22 keV, and its size did not exceed 15 μm in vertical and horizontal directions. The data were acquired with a 2D MAXIPIX detector. The reflectivity data from the curved liquid surfaces were treated according to the procedure described by Konovalov *et al*. [34].

## Declaration of Competing Interests

The authors declare that they have no known competing financial interests or personal relationships that could have appeared to influence the work reported in this paper.

## Data availability

The X-ray data collected during the beamtime experiment MA-5338 at ID10 beamline of the ESRF is available at https://data.esrf.fr/doi/10.15151/ESRF-ES-787695481. The microscopy data can be provided upon reasonable request.

## Author Contributions

The manuscript was written through the contributions of all authors. All authors approved to the final version of the manuscript.

## Acknowledgements

The authors express their gratitude to the European Synchrotron Radiation Facility for providing access to synchrotron radiation facilities and for allowing the use of the surface scattering end-station at beamline ID10. We are grateful to Irina Snigireva (ESRF, Grenoble) for the help with SEM EDX measurements. Additionally, the authors acknowledge the financial support provided by Grant Agreement No. 951943 (DirectSepa) from the European Union's Horizon 2020 research and innovation program.

## Supplementary Material

Supplementary data to this article (alloys' emissivity, quantitative characteristics of flakes, extended table of the XRR fit parameters, alloy composition determination with SEM EDX, and movies of the growths) can be found online at